# Room temperature tunneling anisotropic and collinear magnetoresistance

A.N. Grigorenko and K.S. Novoselov,

*Department of Physics and Astronomy,*

*University of Manchester, Manchester M13 9PL, United Kingdom,*

D.J. Mapps,

*Centre for Research in Information Storage Technology,*

*Department of Communication, Electronic and Electrical Engineering,*

*University of Plymouth, Drake Circus, Plymouth, Devon PL4 8AA, United Kingdom.*

**Abstract**

We report a room temperature tunneling anisotropic magnetoresistance in Co/Al$_2$O$_3$/NiFe junctions containing magnetic electrodes oxidized prior to forming the Al$_2$O$_3$ layer. A significant change in a tunnel magnetoresistance is observed when the layer magnetizations are rotated collinearly in the junction plane by an applied external field. The angular dependence of the tunneling anisotropic magnetoresistance could be explained by the presence of an antiferromagnetic oxide layer formed within the barrier.

PACS: 73.40.Gk, 75.30.Gw, 75.70.-i, 85.30.Mn

Spin-polarized electron tunneling through a thin dielectric barrier attracts attention of many research groups and from different perspectives [1-9]. In magnetic tunnel junctions the tunneling process is significantly complicated by the presence of spin polarization, magnetization dynamics, etc. In a past decade, a set of new effects has been reported and a significant progress has been achieved towards the understanding of spin-polarized tunneling [1-11]. One of such effects consists in unexpected spin valve-like tunnel magnetoresistance observed for Au/Ti/AlOx/(Ga,Mn)As/GaAs structures at low (liquid helium) temperatures [12]. This phenomenon is called tunneling anisotropic magnetoresistance (TAMR) and originates from strong spin-orbital coupling. Soon it was demonstrated that TAMR can be very large in (Ga,Mn)As nanoconstrictions [13] and may open new directions in spintronics applications provided the effect could survive at room temperature.

Recently, we have reported the presence of room temperature TAMR in the Co/$Al_2O_3$/NiFe magnetic tunnel junctions with a magnetic layer oxidized prior to forming the $Al_2O_3$ layer [14] (OMTJ). We have attributed observed TAMR not to spin-orbital coupling but to a thin antiferromagnetic layer of $\alpha$-$Fe_2O_3$ formed at the interface of the oxidized permalloy (NiFe) films [15, 16]. The field dependence of electron tunneling through an antiferromagnetic dielectric barrier offers a qualitative explanation for the measured data [14, 17]. In this paper we report room temperature TAMR observed both in conventional OMTJ and OMTJ with crossed magnetic anisotropies. We demonstrate that magnetoresistance curves of the junctions are strongly dependent on the angle between the applied field and the anisotropy directions. As in Ref. [12], the TMR signal of our junctions changes its sign for some angles of the applied field. We provide a simple model which describes the main properties of the measured TAMR.

The Co/Al$_2$O$_3$/NiFe and Co(CoFe)/Al$_2$O$_3$/NiFe junctions have been fabricated on a silicon substrate in Nordico2000 system and patterned using optical lithography. The physical and structural properties of the films sputtered in our system are discussed in [18]. The anisotropies in the magnetic layers were induced by permanent SmCo magnets placed in the substrate holder during deposition. We have grown junctions with the "conventional" parallel anisotropies, where the anisotropies were induced along the same direction for both layers, and with the "crossed" anisotropies, where the anisotropies were induced in perpendicular directions. The bottom leads were produced by 5nm of Cr followed by 80nm of Au and subsequent plasma etching after optical lithography. After junction lithography we deposited the bottom FeNi magnetic layer and oxidized it in 210 mTorr of pure oxygen for 4 hours inside the chamber. For the conventional magnetic junctions, an Al layer has been then deposited onto bottom magnetic layer and naturally oxidized in 300 mTorr of pure oxygen for 14 hours in the chamber. This followed by deposition of the top Co or CoFe magnetic layer and subsequent lift-off. Hard baked photoresist 1813 (20min at a hot plate of 210°) with necessary lithography was used to isolate the leads. The top lead was produced by sputtering of 5nm of Cr followed by 80nm of Au (the reverse sputtering has been used to clean the top magnetic layer and insure good electrical contact with gold) and subsequent plasma etching after optical lithography. For junctions with crossed anisotropies, after the oxidation of the bottom permalloy layer the sample has been taken out of the chamber and repositioned in the sample holder in order to achieve crossed anisotropies. Usually, the procedure of breaking vacuum leads to a strong deterioration of the magnetic tunnel effect. However, the oxidation of the bottom magnetic layer provided a shield for the layer and the deterioration of the magnetic tunnel effect was less than expected. After sample repositioning, the chamber has been pumped again and an Al layer has been deposited after short reverse sputtering of the bottom magnetic layer. The rest of the fabrication procedure for junctions with crossed anisotropies coincides with that for the conventional ones. The results in this paper are presented for the Co(30nm)/Al$_2$O$_3$(1.25nm)/NiFe(20nm)/Ta(5nm)/Si(substrate) circular junction of 100μm in radius and R$_A$=7Ohm with the parallel anisotropies (junction A), and for the Co(CoFe)(40nm)/Al$_2$O$_3$(1nm)/NiFe(20nm)/Ta(5nm)/Si(substrate) circular junction of 20μm in radius R$_B$=17Ohm with the crossed anisotropies (junction B). Figure 1 shows the schematics of the structure for both junctions described in the paper. The magnetization curves and the layer anisotropies for the larger junctions (radius 500μm) made at the same substrate have been measured in a magnetometer and given in Ref. [19]. The magnetization curve of the conventional junctions showed two distinct switching fields corresponding to two different in-plane uniaxial anisotropies of magnetic layers [19]. We did not measure the magnetization curves and the in-plane uniaxial anisotropies in the smaller junctions directly. These anisotropies (and coupling between layers) have been extracted from the TMR response by micromagnetic calculations [19]. The corresponding anisotropy field values were 4-7Oe in NiFe-layer and 14-20Oe in Co, Co(CoFe) layers and were close to the values obtained from magnetization curves. The intrinsic junction resistance scaled inversely with the junction area and was about 200kOhm×μm$^2$ at low bias for the Co/Al$_2$O$_3$/NiFe structure (17kOhm×μm$^2$ for the Co(CoFe)/Al$_2$O$_3$/NiFe structure with crossed anisotropies) with the maximal TMR value of about 15% (room temperature), which is close to the values of Ref. [20]. The junctions showed similar behavior as far as the angular dependence of TMR is concerned. The exchange bias fields due to an antiferromagnetic layer of oxidized permalloy are supposed to be small at room temperature [22], which was supported by our micromagnetic modeling of TAMR. The square resistance of the leads (defined as the ratio of the gold conductivity over the gold thickness) at the junction region was about 0.5Ohm, which is an order of magnitude smaller that the junction resistances. This suggests that crowding [23] should not affect the TMR measurements. Further details of the junction fabrication and their physical properties are given elsewhere [19].

To demonstrate TAMR in our samples we have measured the tunnel magnetoresistance of *saturated* tunnel junctions in a rotating magnetic field of constant large amplitude produced by a pair of rotating magnets. The junction has been placed in the magnetic field $H_{ext}$ of about $10^3$Oe slowly rotating in the junction plane and the junction resistance has been measured as a function of the angle of rotation (at a fixed magnitude of the applied field). The applied field $H_{ext}$ was well in excess of the layer magnetic anisotropies induced during junction fabrication and was large enough to suppress formation of magnetic domains. At any given orientation the applied field aligned the layer magnetizations along its own direction. The angle of misalignment $\delta\theta_M$ between magnetizations at the applied field of $H_{ext}=10^3$Oe caused by the difference in the layer in-plane anisotropies can be easily evaluated from the micromagnetic theory [19] to be less than $\delta\theta_M \leq 10^{-2}$ rad for any angle φ between the applied field and the anisotropy direction. Conventionally, the angular TMR dependence comes from combining the Julliere's arguments [10] with either the Slonczewski's model of tunneling [11] or the general Landauer-Büttiker formula [24-26] and can be written as [11, 26]

$$\Delta G / G = P_1 P_2 \cos(\theta), \tag{1}$$

where $G$ is the conductance of the tunnel junction, $P_1$ and $P_2$ are spin polarizations and $\theta$ is the angle between the quantization axes (directions of the effective fields). It is also conventionally assumed [11, 24-26] that the angle between quantization axes coincides with the angle between magnetizations. Hence, the relative changes of conductance of the saturated tunnel junction should be less than $\delta G / G < \delta(\cos(\theta)) \approx (\delta \theta_M)^2 / 2 \leq 5 \cdot 10^{-5}$ in the case of magnetizations following the rotation of the applied magnetic field of 1kOe. Figure 2 demonstrates that this is not the case experimentally.

Figure 2 shows the change in the resistance of the junction A (squares) and the junction B (circles) as a function of the in-plane angle of the applied field $H_{ext}$ which magnitude was kept constant at $10^3$Oe. (The change is normalized by the sample resistance and the angle of the external field is taken with respect to the direction of the induced anisotropy). The observed change (16%) in the tunnel resistance of the saturated junction for the junction A is of the order of the resistance change (14%) observed in the conventional TMR dependence. The variation of the resistance of the saturated junction *relative to the conventional TMR change* is about 120% and five orders of magnitude bigger than the value $(\delta \theta_M)^2 / 2 \leq 5 \cdot 10^{-5}$ expected by the conventional angular TMR dependence (1). This striking feature of the angular dependence of the saturated tunnel junctions, necessarily implies that TMR is not simply determined by the angle between magnetizations but depend on the direction of magnetizations with respect to the (induced) crystalline anisotropy, which is the characteristic feature of TAMR [12, 13].

The collinear TMR (which appears when the layer magnetizations rotate collinear in the junction plane) has been observed for all studied OMTJ irrespectively of their geometrical form (circle, square, rhomb) and size (down to 2μm junctions), which rules out the shape anisotropy as the source of the phenomenon. The current distribution inside the junction ("current crowding" [23]) should not change for the collinear rotation of magnetizations and therefore cannot be responsible for the measured TMR. In addition, as was mentioned above, the crowing should be small in our junctions. Directions of the magnetic field at which the minimum resistance was observed in the junction A ($\delta$=0° and 180°) coincide with the directions of the in-plane magnetic anisotropy of both electrodes induced during the junction fabrication. For the junction B with crossed anisotropies the minima are observed at intermediate angles of $\delta$=43° and 223°. The collinear TMR for the junction A (and B) is described well by the $\cos^2(\varphi+\delta)$ function shown as the solid line in Fig. 2. These features suggest that the induced magnetic anisotropies are responsible for the collinear TMR.

To gain further information about this unexpected phenomenon, we have measured the junction magnetoresistance as a function of the magnetizing field for a variety of field orientations with respect to the anisotropy directions. The main feature of TAMR dependences is the change of the sign of the TMR effect for some orientations of the magnetizing fields. The positive TMR observed in the junction A magnetized by the field applied along the direction of the magnetic anisotropies changes to the negative TMR observed in the same junction for the magnetic field applied at the angle 70° with respect to the anisotropy direction (Fig. 3(a)). Figure 4 shows a set of TAMR dependences with positive and negative TMR measured at different angles of the applied field for OMTJ with crossed anisotropies (junction B).

We found that the measured TAMR (as well as the collinear TMR) is described well by the micromagnetic model [14, 19]. In this model the magnetization response is calculated through the pair of coupled Landau-Lifshits equations for two magnetic layers [19], while the TMR effect is calculated using the theory developed in [14]. According to the theory the conductivity $G_f$ of the junction with ferromagnetic barrier is given by

$$G_f = G_0 \left( \left(1 + P_1 P_2 \cos(\theta)\right) \cosh(k_{eff} d) + \left(P_1 \cos(\chi) + P_2 \cos(\varphi)\right) \sinh(k_{eff} d) \right), \quad (2)$$

where $k_{eff} = J\sqrt{2m}/(\hbar\sqrt{\phi})$ (here $J$ is the exchange constant of the barrier, $m$ is the electron mass, $\phi$ is the barrier height), $\theta$ is the angle between quantization axes of electrodes, $\chi$ is the angle between the quantization axes of the first electrode and the barrier, and $\varphi$ is angle between the quantization axes of the second electrode and the barrier. Here the term proportional to $\cosh(k_{eff} d)$ describes the conventional TMR dependence while the term proportional to $\sinh(k_{eff} d)$ yields the contribution of the ferromagnetic barrier. In the case of an antiferromagnetic barrier (formed in OMTJs by a thin layer of α-Fe$_2$O$_3$ at the interface of the oxidized permalloy film [15, 16]) this theory should be modified. In the first approximation, the $\cos(\chi)$ and $\cos(\varphi)$ dependence (which appears due to projection of the electrode magnetization **M** onto the barrier antiferromagnetic vector **L**) should be replaced by the $\cos^2(\chi)$ and $\cos^2(\varphi)$ due to symmetry considerations (the scalar product of magnetization with an antiferromagnetic vector $\mathbf{M} \cdot \mathbf{L}$ is an axial quantity while the conductivity is a polar quantity), see Ref. [27]. Therefore, the conductivity $G_{af}$ of the junction with antiferromagnetic barrier can be written approximately as

$$G_{af} = G_0\left((1+P_1P_2\cos(\theta))\cosh(k_{eff}d) + \left(P_1\cos^2(\chi) + P_2\cos^2(\varphi)\right)\sinh(k_{eff}d)\right). \quad (3)$$

Using (3), we have simulated the observed TMR dependences in Figs. 3, 4 by choosing the appropriate values of the in-plane anisotropies and exchange constants for the two magnetic layers and the antiferromagnetic barrier [14]. For example, Fig. 3(b) shows the calculated TMR response of the junction A to the magnetizing field applied at the angle of 70° with respect to the magnetic anisotropies (the corresponding experimental curve is given in Fig. 3(a)). Parameters for calculations were: saturation magnetizations $4\pi M_{NiFe}$=0.8T, $4\pi M_{Co}$=1.8T, the Landau-Lifshits constants $\lambda_{NiFe}$=2770G$^{-2}$s$^{-1}$, $\lambda_{Co}$=1220G$^{-2}$s$^{-1}$, spin polarizations $P_1$=0.4, $P_2$=0.8, induced uniaxial in-plane anisotropies $H_{NiFe}$=5Oe, $H_{Co}$=15Oe and $th(k_{eff}d)$=1.4. Figure 4(c) provides the result of micromagnetic simulations for the junction B with magnetic field applied at the angle 110° with respect to the magnetic anisotropy of the bottom electrode (the experimental curve is shown in Fig. 4(b)). The parameters for micromagnetic calculations were: $4\pi M_{NiFe}$=0.8T, $4\pi M_{CoFe}$=2T, the Landau-Lifshits constants $\lambda_{NiFe}$=2770G$^{-2}$s$^{-1}$, $\lambda_{CoFe}$=1100G$^{-2}$s$^{-1}$, spin polarizations $P_1$=0.4, $P_2$=1, induced uniaxial in-plane anisotropies $H_{NiFe}$=1.1Oe, $H_{Co}$=14Oe and $th(k_{eff}d)$=3. One can see a good agreement between the result of modeling and the measured TAMR. In addition, the formula (3) provides an excellent description of the collinear TMR shown in Fig. 2.

In conclusion, we have observed room temperature tunneling anisotropic magnetoresistance in Co/Al$_2$O$_3$/NiFe junctions containing magnetic electrodes oxidized prior to forming the $Al_2O_3$ layer and observed unexpectedly large changes of TMR in the saturated junctions where layer magnetizations (being collinear) are rotated with respect to the induced in-plane magnetic anisotropies.

Authors thank Andre Geim for useful discussions, Nick Fry and Phil Brown for technical assistance.

**References.**

**Figure Captions.**

Fig. 1. The schematic structure of the junctions A and B. The arrows on magnetic layers show the direction of induced anisotropies. The optical photographs of the junctions A and B are given on the right.

Fig. 2.

TAMR as a function of the angle φ between the external field $H_{ext}$ of a fixed amplitude ($10^3$ Oe) and the axis of induced layer anisotropies: squares – the junction A, circles – the junction B. The inset shows schematically the experimental geometry (the circle corresponds to the junction).

Fig. 3.

(a) Measured TMR as a function of the magnitude of the in-plane field $H_{ext}$ applied at an angle of 70° to the direction of magnetic anisotropies of the junction A; the inset shows schematically the experimental geometry. (b) TMR calculated within the micromagnetic theory using expression (3) at the same conditions as in (a).

Fig. 4.

(a)-(b) TMR measured at different angles φ of the in-plane field $H_{ext}$ applied to the junction B. The inset shows schematically the experimental geometry. (c) TMR calculated within the micromagnetic theory using expression (3) for the angle of the applied in-plane field of φ=110°.

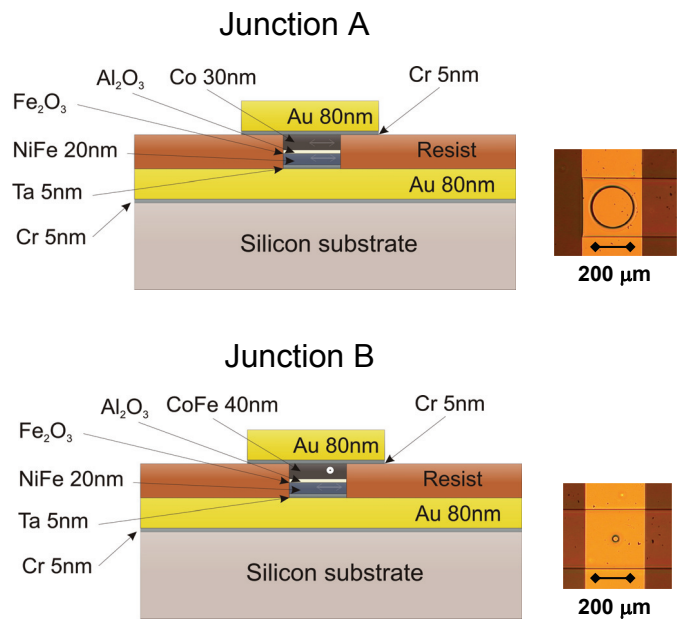

Fig. 1

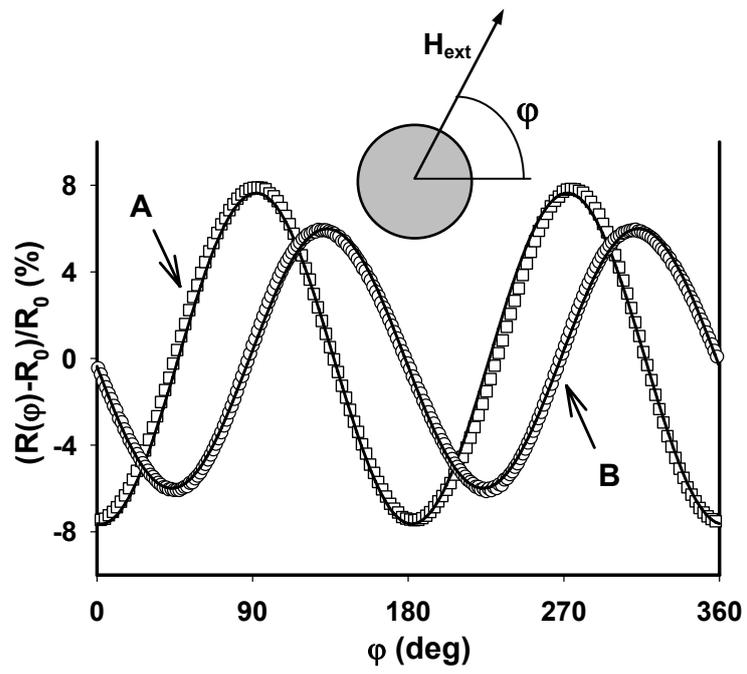

Fig. 2

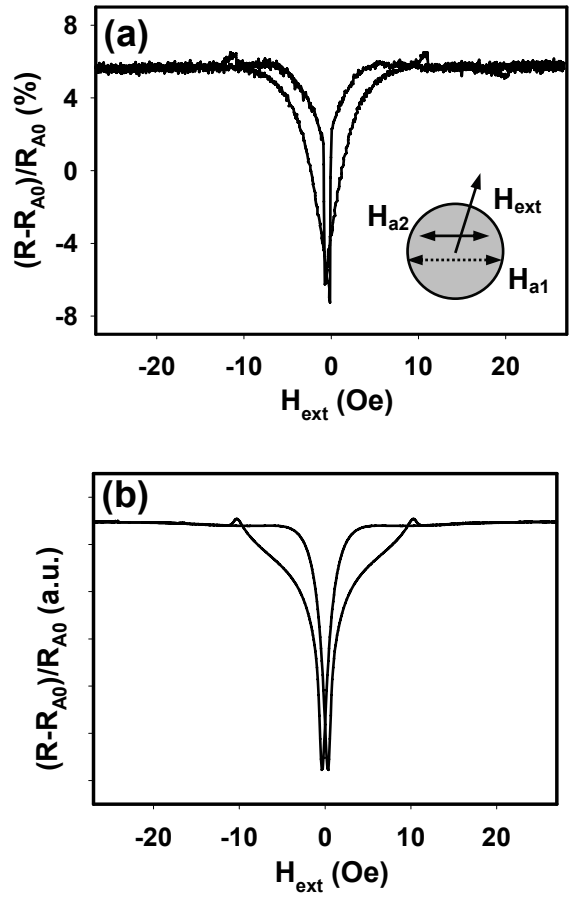

Fig. 3

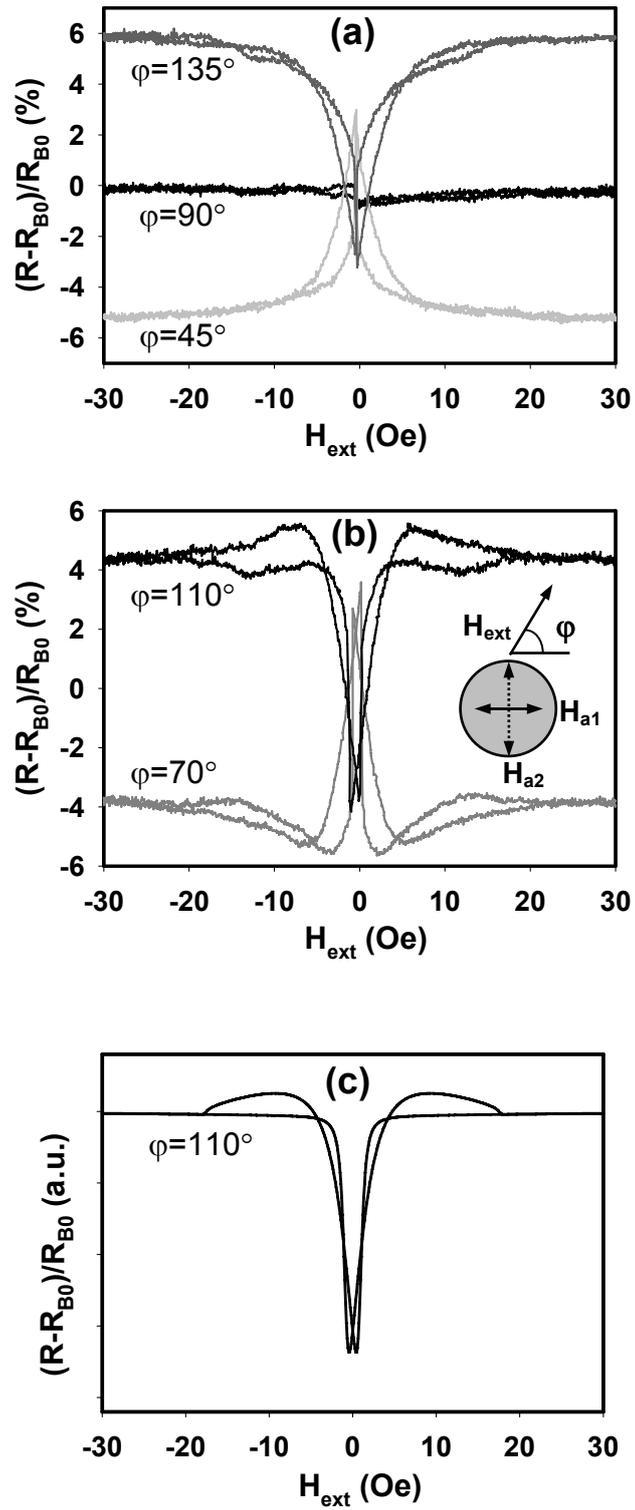

Fig. 4